\documentclass[12pt]{iopart}
\usepackage{citesort}
\usepackage{graphicx}
\expandafter\let\csname equation*\endcsname\relax
\expandafter\let\csname endequation*\endcsname\relax
\usepackage{amsmath,amssymb}

\usepackage{ctable}

\newcommand{\pref}[1]{(\ref{#1})}
\newcommand*{\bsb}{\boldsymbol}

\begin{document}
\title{Optomechanical interactions in non-Hermitian photonic molecules}

\author{D~W~Sch{\"o}nleber$^1$, A~Eisfeld$^1$ and R~El-Ganainy$^{1,2}$}
\address{$^1$ Max Planck Institute for the Physics of Complex Systems, N\"othnitzer Strasse 38, 01187 Dresden, Germany}
\address{$^2$ Department of Physics and Henes Center for Quantum Phenomena, Michigan Technological University, Houghton, Michigan, 49931, USA}

\begin{abstract}
We study optomechanical interactions in non-Hermitian photonic molecules that support two photonic states and one acoustic mode. The nonlinear steady-state solutions and their linear stability landscapes are investigated as a function of the system's parameters and excitation power levels. We also examine the temporal evolution of the system and uncover different regimes of nonlinear dynamics. Our analysis reveals several important results: (1) Parity-time ($\mathcal{PT}$) symmetry is not necessarily the optimum choice for maximum optomechanical interaction. (2) Stable steady-state solutions are not always reached under continuous wave (CW) optical excitations. (3) Accounting for gain saturation effects can regulate the behavior of the otherwise unbounded oscillation amplitudes. Our study provides a deeper insight into the interplay between optical non-Hermiticity and optomechanical coupling and can thus pave the way for new device applications. 
\end{abstract}

\maketitle

\section{Introduction}\label{sec:intro}
Cavity optomechanics (COM) has attracted considerable attention on both theoretical and experimental fronts during the past decade \cite{kippenberg2008,aspelmeyer2014a,aspelmeyer2014,brennecke2008}. This was largely enabled by the rapid increase of computational power that allowed for accurate simulations of optomechanical coupling, and the recent progress in fabrication and measurement techniques that led to experimental observation of this interaction in different material setups. Nowadays, optomechanical interactions are being utilized in various applications such as gravitational wave detectors \cite{arcizet2006,mcclelland2011}, quantum memories \cite{chang2011} and acceleration sensors \cite{krause2012}, just to mention a few. Furthermore, optical cooling of macroscopic mechanical oscillators \cite{sawadsky2015} provides a unique opportunity to study the classical-quantum correspondence.

A different notion that has gained a lot of attention recently is parity-time ($\mathcal{PT}$) symmetry where it was shown that certain $\mathcal{PT}$ symmetric Hamiltonians can posses real spectra \cite{bender1998}. This concept was later extended to optics \cite{el-ganainy2007,makris2008,musslimani2008,guo2009}, where its experimental manifestations were observed for the first time in optical systems with engineered gain and loss profiles \cite{ruter2010}, as well as other fields (see, e.g., \cite{zhang2015}). Noteworthy, most of the intriguing features of $\mathcal{PT}$ symmetric structure also persist for the wider class of non-Hermitian material that do not necessarily respect $\mathcal{PT}$ symmetry. For example, the existence of the spectral singularities known as exceptional points (EPs) do not require $\mathcal{PT}$ symmetry and can occur in a general non-Hermitian system \cite{heiss2004,mueller2008,el-ganainy2014}. The ability to manipulate light in photonic systems by controlling these singularities has opened the door for new device applications such as single mode microring lasers \cite{feng2014,hodaei2014} and light sources based on non-Hermitian phase matching \cite{el-ganainy2015}.

Recently, the marriage between the two themes of optomechanics and $\mathcal{PT}$ symmetry has been proposed \cite{jing2014}. In particular, this pioneering work has investigated the optomechanical coupling in a phonon laser (or saser) structures similar to those studied in Ref.~\cite{grudinin2010} but with the additional ingredient of $\mathcal{PT}$ symmetry. The analysis in Ref.~\cite{jing2014} predicted in particular a giant enhancement of the optomechanical coupling strength around the exceptional points. However, not all important questions regarding the stability and dynamics of the system have been answered in this work.

In order to fill in this gap, we perform a comprehensive analytical and numerical investigation of optomechanical interactions in non-Hermitian photonic molecules similar to those considered in \cite{jing2014}. We  characterize the nonlinear steady-state solutions and the stability properties in terms of the optical and acoustic design parameters as well as the optical pumping power levels and frequencies.
Furthermore, we study the dynamical evolution of the system and show that different regimes of operations can be identified based on the design parameters, excitation power levels and the frequency detuning. Our study reveals several important results: (1) The maximum achievable optomechanical interaction enhancement for stable steady-state solutions does not occur in the neighborhood of the $\mathcal{PT}$ phase transition point, and (2) Depending on the design parameters, pump properties and gain saturation effects, different regimes of nonlinear dynamics such as fixed points and sustained oscillations are possible.

\section{System and model}\label{sec:system_model}
\begin{figure}[tb]
\centering
\includegraphics[width=.75\columnwidth]{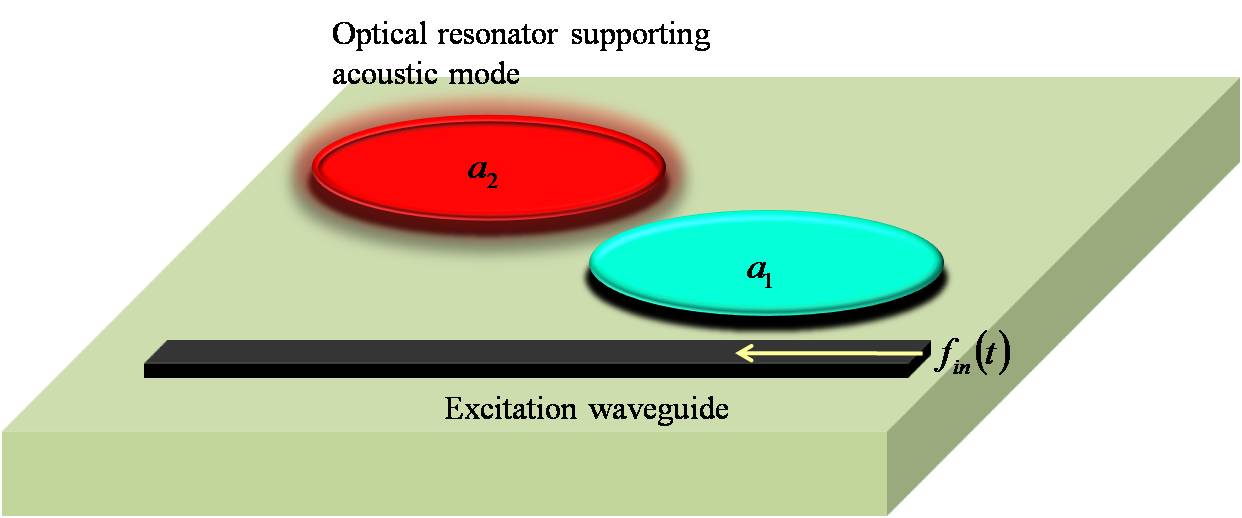}
\caption{A schematic of the optomechanical system under consideration. It consists of two coupled optical resonators $a_{1,2}$, each having a finite quality factor and experiencing optical gain or loss due to optical or electrical pumping (not shown here). The gain/loss profile across the cavities is in general asymmetric as indicated by their different colors.  The halo surrounding resonator $a_2$ indicates that it supports a vibrational mechanical mode at frequency $\omega_m$. Optical excitation of the system takes place via the evanescent coupling between resonator $a_1$ and an external waveguide.}
\label{fig:Fig1_system}
\end{figure}

In this work we consider a photonic molecule made of two optically identical resonators that supports two photonic supermodes and one acoustic mode as shown schematically in \Fref{fig:Fig1_system}. Similar to the work in \cite{grudinin2010,jing2014,guo2014,wang2014,lue2015,jing2015,liu2015}, the acoustic mode is assumed to be localized only in one of these cavities and is characterized by a resonant frequency $\omega_m$, damping coefficient $\Gamma$ and an effective mass $m$, while the optomechanical coupling is characterized by the coupling constant $g$. The uncoupled photonic states of the two resonators have identical resonant frequencies $\omega_0$ and quality factors (not necessarily the same) quantified by the inverse of the radiation loss coefficients $\alpha_{1,2}$. The optical coupling coefficient between the two resonators is given by $J$. Furthermore, additional gain or loss factors $\tilde{\gamma}_{1,2}$ can be engineered by an appropriate design of the material system (for instance by doping the resonator with gain/loss material and applying different optical pumping conditions) \cite{jing2014}. The optical excitation is achieved through a waveguide coupled to the acoustically active cavity with a coupling constant $\mu$. As a result, the total net gain/loss in each resonator is described by the coefficients $\gamma_1=\tilde{\gamma}_1-\alpha_1-\mu$ and $\gamma_2=\tilde{\gamma}_2-\alpha_2$. Note that these values can be either positive or negative depending on whether the net effect is optical amplification or decay. In our study, we do not discuss in detail how these gain/loss parameters can be controlled (see \cite{jing2014} for more details on that subject) but rather focus on how their values affect the dynamics. 

Under these conditions and by neglecting quantum correlations and fluctuations, the equations of motion for the (complex) classical optical field amplitudes $a_{1,2}$ in the two resonators and the (real) acoustic oscillator displacement $x$ respect the following nonlinear system of differential equations \cite{aspelmeyer2014,fan2003,peng2014,peng2014a,jing2014,lue2015}:
\begin{subequations}\label{eq:semiclassical_Langevin_eq}
 \begin{align}
  \dot{a}_1 &= (-i\Delta + \gamma_1) a_1 - iJ a_2 + \sqrt{2\mu}f_0,\label{eq:semiclassical_Langevin_eq_a1}\\
  \dot{a}_2 &= (-i\Delta + \gamma_2) a_2 - iJ a_1 -ig a_2 x,\label{eq:semiclassical_Langevin_eq_a2}\\
  \ddot{x} &= -\Gamma \dot{x} -\omega_m^2 x + \frac{\hbar g}{m} |a_2|^2.\label{eq:semiclassical_Langevin_eq_x}
 \end{align}
\end{subequations}
Here, $\Delta$ denotes the laser detuning $\Delta=\omega_0-\omega_L$, and $\mu$ is the coupling rate between the waveguide and the resonator $a_1$. 

Note that the above equations are written in the rotating frame of reference of the optical excitation signal $f_\mathrm{in}(t)=f_0 \exp(-i\omega_L t)$, where $f_0$ is the amplitude of the external excitation laser and $\omega_L$ is its frequency.
The power $P_\mathrm{in}$ of the excitation laser transmitted to the resonator $a_1$ can be obtained from $f_0$ via $P_\mathrm{in}=\hbar\omega_L|f_0|^2$ \cite{lue2015}.

In the absence of any non-Hermiticity, the above system was reported to operate as a saser (acoustic laser) device where the frequency splitting between the photonic supermodes of the photonic molecule can be treated as a two-level system that can provide acoustic gain for the mechanical mode \cite{grudinin2010}.

In what follows we do not emphasize the saser action picture presented in Refs.~\cite{grudinin2010,jing2014} but rather treat the system from the dynamical point of view. Note that we use the physical parameters summarized in \Tref{tab:parameters}. 

\ctable[
  caption = {List of the design parameters that we use throughout the manuscript (cf.\ Refs.~\cite{grudinin2010,peng2014,jing2014,lue2015}) for the numerical calculations.},
  pos = tb, 
  label = tab:parameters,
  captionskip  =  1ex,
  left, 
  botcap, 
  mincapwidth = 15cm, 
  ]{cc}{
}{\FL 
Parameter & Value\ML 
$\omega_m$ & $23.4\times2\pi$ MHz \NN 
$\omega_0$ & $193\times2\pi$ THz \\ & (corresponds to $\lambda_0 = 1.55$ $\mu$m)\NN
$g$ & $5.61$ GHz/nm \NN
$m$ &  $5\times10^{-11}$ kg \NN
$\Gamma$ & $0.24$ MHz \NN
$J$ & $6.45$ MHz \NN
$\gamma_0$ & $6.45$ MHz \NN
$\mu$ & $3.14$ MHz \LL 
}

\section{Steady-state solutions and their stability properties}\label{sec:ss_stability}
\subsection{Steady-state analysis}\label{subsec:ss_analysis}
We start our analysis by investigating the steady-state solutions associated with the non-Hermitian optomechanical interaction of the system depicted in \Fref{fig:Fig1_system}. We do so by setting the time derivatives of $a_{1,2}$ and $x$ to zero and solving Eqs.~\pref{eq:semiclassical_Langevin_eq} for the steady state $x_\mathrm{s}$ of the mechanical oscillator. This yields an algebraic cubic polynomial equation that, in general, has three (possibly complex) different solutions. In the following, we only discuss real solutions of $x_\mathrm{s}$, which correspond to a physical oscillator displacement.

To assess the figure of merit for the optomechanical interaction in our system, we first consider a reference system with both optical resonators having identical losses, i.e., $\gamma_1=\gamma_2<0$, before analyzing the full non-Hermitian system with optical gain. The resulting mechanical steady-state amplitude $x_\mathrm{s,p}$ in that latter case serves as a reference to estimate the enhancement $\eta$ of the system:
\begin{equation}\label{eq:eta_definition}
 \eta = \frac{x_\mathrm{s}}{x_\mathrm{s,p}}.
\end{equation}
Here, the subscript $\mathrm{p}$ in Eq.~\pref{eq:eta_definition} denotes the passive case, i.e., the case where both resonators have losses.

It is worth noting that in the work by Jing \emph{et al.} \cite{jing2014}, a strong enhancement $\eta$ of two orders of magnitude has been found when $\gamma_2=-\gamma_1$ under resonant excitation conditions, i.e. $\Delta=0$. Here we also explore the case of off-resonant $\Delta\neq0$ driving.
Besides, we note that that in Ref.~\cite{jing2014} a different scaling of the optical amplitude $f_0$ has been used, i.e. $\sqrt{2\gamma_1}$ instead of $\sqrt{2\mu}$. Therefore, the enhancement values found in \cite{jing2014} are scaled with respect to the ones obtained in this manuscript.

\subsubsection{Analytical considerations:}
In order to gain an insight into the behavior of the system beyond the full numerical solution of Eqs.~\pref{eq:semiclassical_Langevin_eq}, we first consider the optical modes only and ignore the driving term while accounting for the nonlinear interaction between the mechanical oscillator and optical amplitude $a_2$ through a non-linearly induced frequency shift. In other words, we treat the steady-state displacement of the mechanical oscillator $x_\mathrm{s}$ as a parameter that effectively introduces an additional detuning $\Delta_x \equiv g x_\mathrm{s}$ to the second cavity. Note that this detuning in reality depends on the strength of the laser driving; a feature that is absent in this simplified analysis. Within this picture, the optical amplitudes are modeled by the following linear equations:
 \begin{equation}\label{eq:semiclassical_Langevin_eq_illustration}
  \partial_t \begin{pmatrix} a_1 \\ a_2  \end{pmatrix} = 
  -i \begin{pmatrix}
    i \gamma_1 & J\\
    J & \Delta_x + i\gamma_2
  \end{pmatrix}
  \begin{pmatrix} a_1 \\ a_2  \end{pmatrix}.
 \end{equation}
By diagonalizing Eqs.~\pref{eq:semiclassical_Langevin_eq_illustration}, we obtain the eigenfrequencies of the two supermodes as well as the associated linewidths as given by the real and imaginary parts, respectively, of the complex frequencies
\begin{equation}\label{eq:evals_illustrative_eq}
 \omega_\pm = \frac{1}{2}\left(\Delta_x + i(\gamma_1+\gamma_2) \pm \sqrt{4J^2 + (\Delta_x-i(\gamma_1-\gamma_2))^2} \right).
\end{equation}
Hence, by scanning the frequency of the pump laser [represented by $\Delta$ in Eqs.~\pref{eq:semiclassical_Langevin_eq}] to match the real part of either $\omega_\pm$, resonant interaction is expected to take place. From \Eref{eq:evals_illustrative_eq}, the following important features can be observed:
\begin{itemize}
 \item[(i)] For antisymmetric gain/loss profile  ($\gamma_2=-\gamma_1<0$) and low laser power ($\Delta_x \ll J, \gamma_1$), we expect the system to exhibit two sharp resonances at $\pm \sqrt{J^2-\gamma_1^2}$ for $J>\gamma_1$, while for $J<\gamma_1$ we expect a single broad resonance at zero.
 \item[(ii)] For smaller values of the gain coefficient, $0<\gamma_1<|\gamma_2|$ with $\gamma_2<0$ and nonzero $\Delta_x$, the square root in Eq.~\pref{eq:evals_illustrative_eq} has both real and an imaginary parts, which we denote by $\Re $ and $\Im$, respectively. In this regime, the resonance frequencies of the supermodes and their associated linewidths are given by $\Delta_x/2 \pm \Re /2$ and $(\gamma_1+\gamma_2)/2 \pm \Im/2$, correspondingly and we expect an asymmetric spectrum for positive and negative laser detuning, respectively.
\end{itemize}

Our discussion so far has focused on the eigenfrequencies of the optical supermodes of the photonic molecule in the absence of pumping. In order to gain more insight into the system's behavior, we now consider the effect of the driving field in our simple picture, i.e., we add $(\sqrt{2\mu}f_0,0)^T$ to the right hand side of Eq.~\pref{eq:semiclassical_Langevin_eq_illustration}, where the superscript $T$ denote matrix transpose. Under these conditions and by assuming a constant detuning $\Delta_x$, we find that Eqs.~\pref{eq:semiclassical_Langevin_eq_illustration} admits a non-trivial steady-state solution for the field amplitudes $a_{1,2}$. By noting that $x_\mathrm{s} \propto |a_2|^2$ under steady-state conditions [cf.\ Eq.~\pref{eq:semiclassical_Langevin_eq_x}], we find that the efficiency $\eta$ is given by $\eta = |a_2|^2 / |a_{2,\mathrm{p}}|^2$ with $\mathrm{p}$ again indicating the passive case with $\gamma_1=\gamma_2<0$. By evaluating the quantity $|a_2|^2 / |a_{2,\mathrm{p}}|^2$ exactly at the onset of the linear $\mathcal{PT}$ phase transition point, we obtain (see \ref{sec:appendix} for the general case):
\begin{equation}\label{eq:eta_estimate_illustration}
 \eta  = 1+\frac{4\gamma_2^2}{\Delta_x^2} \quad\quad (\gamma_2=-\gamma_1=-J<0).
\end{equation}

Formula \pref{eq:eta_estimate_illustration} indicates that a larger nonlinear-induced detuning $\Delta_x$ will decrease the enhancement factor $\eta$. Since the mechanical amplitude $x_\mathrm{s}$ is proportional to the detuning $\Delta_x$ and increases with laser power, we expect the efficiency $\eta$ to drop as the driving power increases. Note that the reason we explicitly consider the $\mathcal{PT}$ point is that at this point the ratio $|a_2|^2 / |a_{2,\mathrm{p}}|^2$ becomes particularly simple. For more details we refer to \ref{sec:appendix}.

Having gained some qualitative inside into the problem by using this simplified linearized analysis, we now turn to the discussion of the numerical steady-state results of the full nonlinear system of Eqs.~\pref{eq:semiclassical_Langevin_eq}.

\subsubsection{Numerical evaluation of the steady-state solutions:}

\begin{figure}[tb]
\centering
\includegraphics[width=.8\columnwidth]{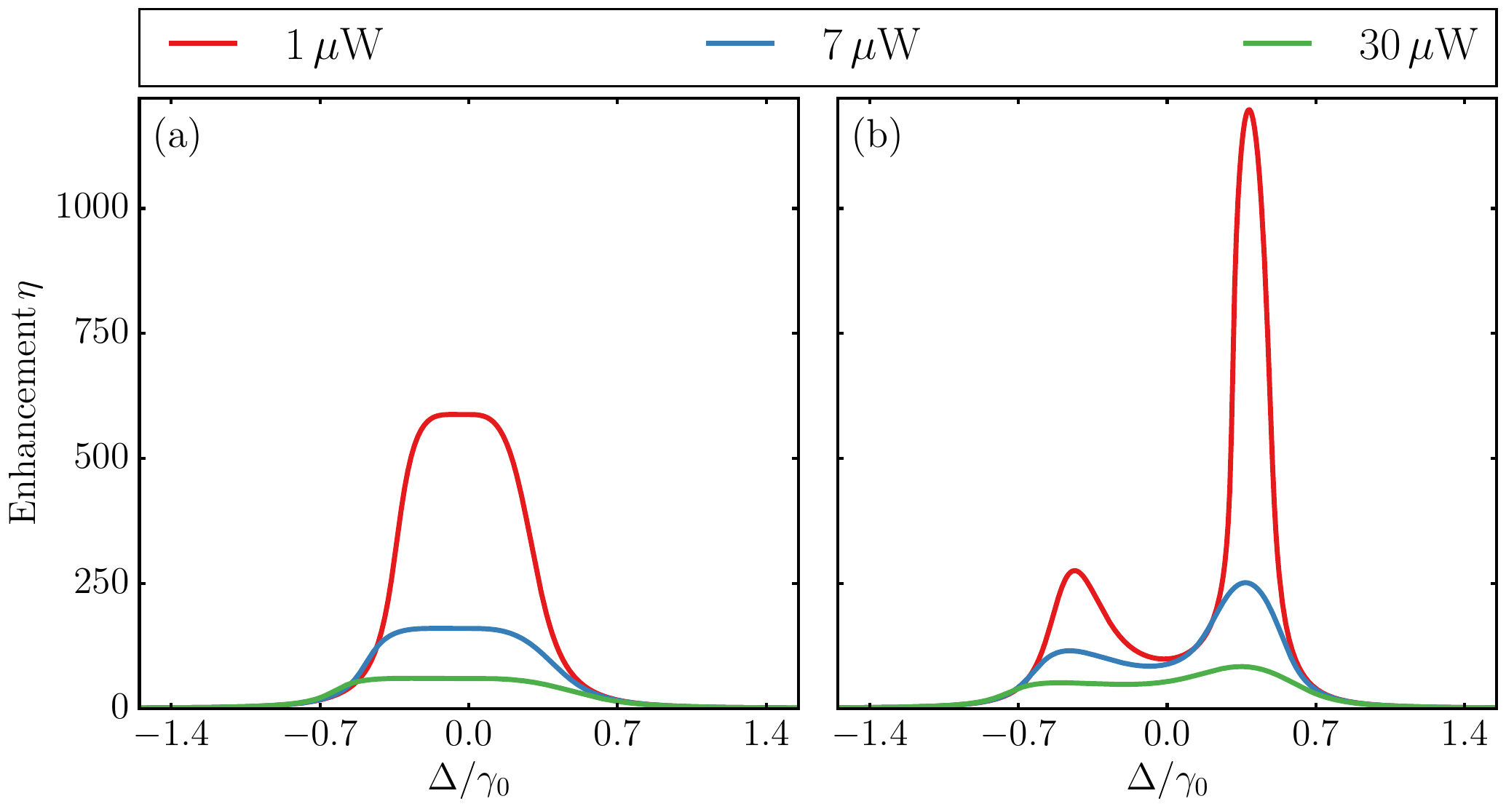}
\caption{Enhancement $\eta$ as a function of the detuning $\Delta$ for various laser powers $P_\mathrm{in}$. In (a), $\gamma_1=\gamma_0$ and $\gamma_2=-\gamma_0$ whereas in (b) $\gamma_1=0.8\gamma_0$ and $\gamma_2=-\gamma_0$. The red, blue, and green lines (from top to bottom) correspond to laser powers of $P_\mathrm{in}=1\mu$W, $7\mu$W, and $30\mu$W, respectively. Other parameters as in \Tref{tab:parameters}.}
\label{fig:Fig2_eta_kappasmall_deltascan}
\end{figure}

We now consider the full numerical evaluation of the steady-state solutions of Eqs.~\pref{eq:semiclassical_Langevin_eq} under general conditions. \Fref{fig:Fig2_eta_kappasmall_deltascan}(a) shows the enhancement factor $\eta$ as a function of $\Delta/\gamma_0$ in the $\mathcal{PT}$ symmetric case where $\gamma_1=-\gamma_2=\gamma_0$. In this scenario, the enhancement curve displays a plateau with no sharp peaks and its maximal value is found to occur at zero detuning $\Delta=0$. Note that the point of maximal enhancement ($\Delta = 0$ and $J=\gamma_1=-\gamma_2$) coincides with the exceptional point, at which the eigenfrequencies of the supermodes of the linear system coalesce.

In \Fref{fig:Fig2_eta_kappasmall_deltascan}(b), the case of unbalanced gain/loss profile, $\gamma_1<\gamma_0$ and $\gamma_2=-\gamma_0$, is shown. In this case, two peaks of different heights that correspond to two different laser detunings can be observed in the enhancement curve. Notably, at the location of the positive detuning peak, the enhancement value even exceeds the one found for the $\mathcal{PT}$ symmetric case at resonance. Our analysis thus uncovers the important result that $\mathcal{PT}$ symmetry is not necessarily the optimum choice for obtaining stronger optomechanical interactions as compared to a passive system. 
These results clearly show that the enhancement of the optomechanical coupling coefficient is not simply an outcome of increasing the optical gain in resonator $a_1$, but rather a result of a complex interplay between the non-Hermitian parameters of the system (optical gain and loss), detuning between the pump laser and the resonance frequency of the optical cavities as well as the properties of the acoustic mode.
The asymmetry observed for the broken $\mathcal{PT}$ symmetry case [\Fref{fig:Fig2_eta_kappasmall_deltascan}(b)] can be understood in the light of our simplified picture of the previous section where the effective detuning introduced to the second cavity $a_2$ due to optomechanical interaction was shown to introduce an asymmetry to the supermode frequencies and linewidths.

Surprisingly, as the driving laser power is increased, the enhancement values drop, indicating that the difference in the mechanical steady-state displacement between the active-passive (gain/loss) and passive-passive (loss/loss) system vanishes. This feature is consistent with our simplified picture introduced in the previous section where the nonlinearly-induced detuning $\Delta_x$  was shown to degrade the enhancement factor [see Eq.~\pref{eq:eta_estimate_illustration}].

\begin{figure}[tb]
\centering
\includegraphics[width=.8\columnwidth]{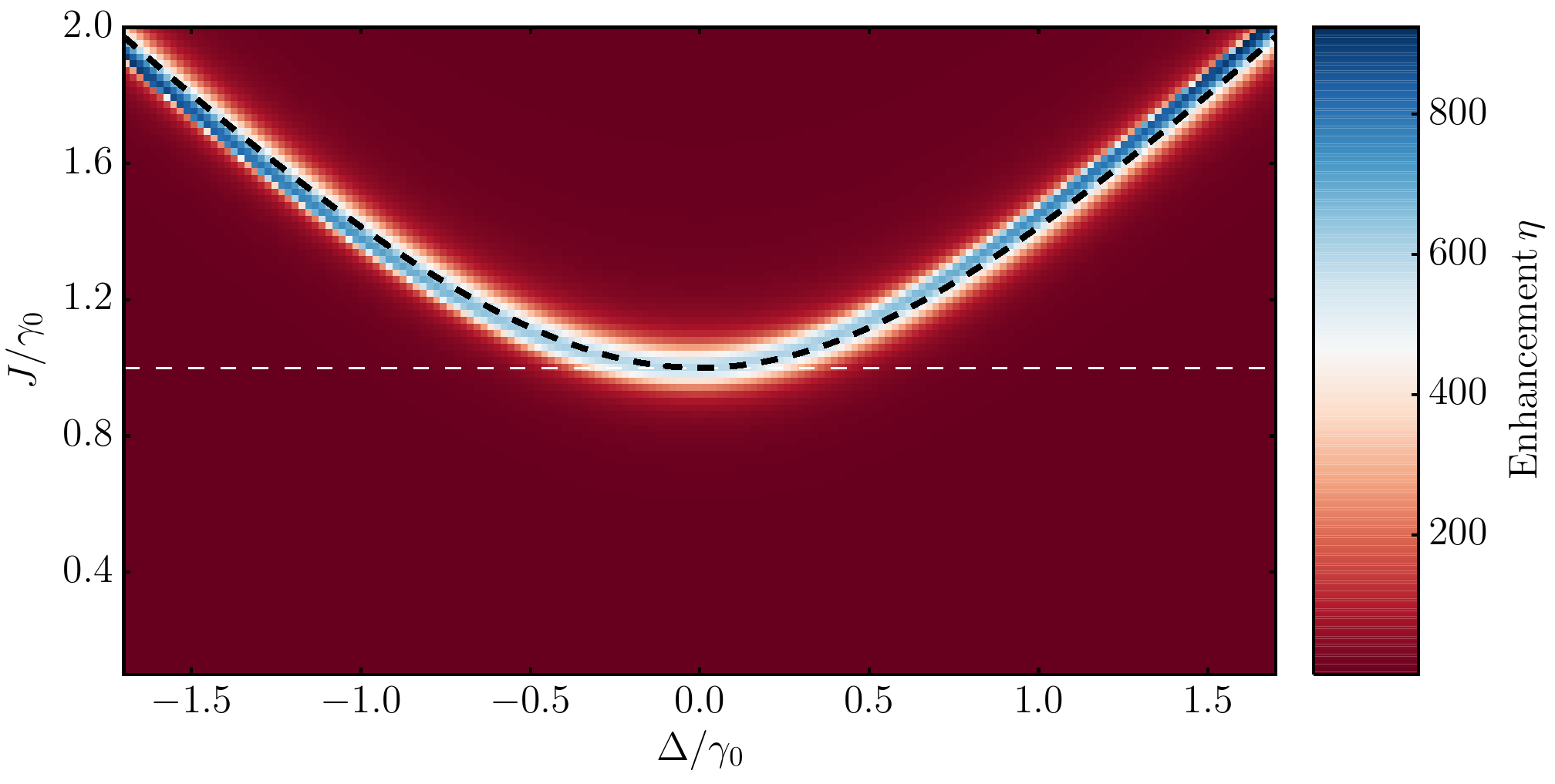}
\caption{Enhancement $\eta$ as a function of the detuning $\Delta$ and the optical inter-cavity coupling $J$ at a driving power of $P_\mathrm{in}=1$ $\mu$W and $\gamma_1=-\gamma_2=\gamma_0$. The horizontal white dashed line is a cross section corresponding to the parameters of \Fref{fig:Fig2_eta_kappasmall_deltascan}(a). The dashed black line shows the $\sqrt{J^2-\gamma_0^2}$ dependence of the eigenfrequencies found from the simplified picture in the previous section [Eq.~\pref{eq:evals_illustrative_eq}]. Other parameters as in \Tref{tab:parameters}.}
\label{fig:Fig3_eta_J_deltascan}
\end{figure}

The resonant behavior of the enhancement curve [the appearance of two sharp peaks in Fig.~\ref{fig:Fig2_eta_kappasmall_deltascan}(b)] at nonzero detuning can occur not only when the optical gain in one cavity is unequal to the loss in the other, but also in the case of equal gain and loss, provided that the inter-cavity coupling exceeds the gain and loss values. This behavior is illustrated in \Fref{fig:Fig3_eta_J_deltascan}, where the optical inter-cavity coupling $J$ is varied while maintaining balanced gain and loss, $\gamma_1=-\gamma_2=\gamma_0$. For $J>\gamma_0$, the maximal enhancement is no longer found at the excitation resonance $\Delta=0$ but rather shifts to $\Delta \neq 0$, in good agreement with our earlier discussion as outlined in (i) in the previous section (black dashed line in Fig.~\ref{fig:Fig3_eta_J_deltascan}). That is, the sizable enhancement value at the exceptional point (found on the dashed white line for $\Delta=0$) is outperformed by the enhancement obtained for $J>\gamma_0$ at the position of the supermode frequencies.

\begin{figure}[tb]
\centering
\includegraphics[width=.8\columnwidth]{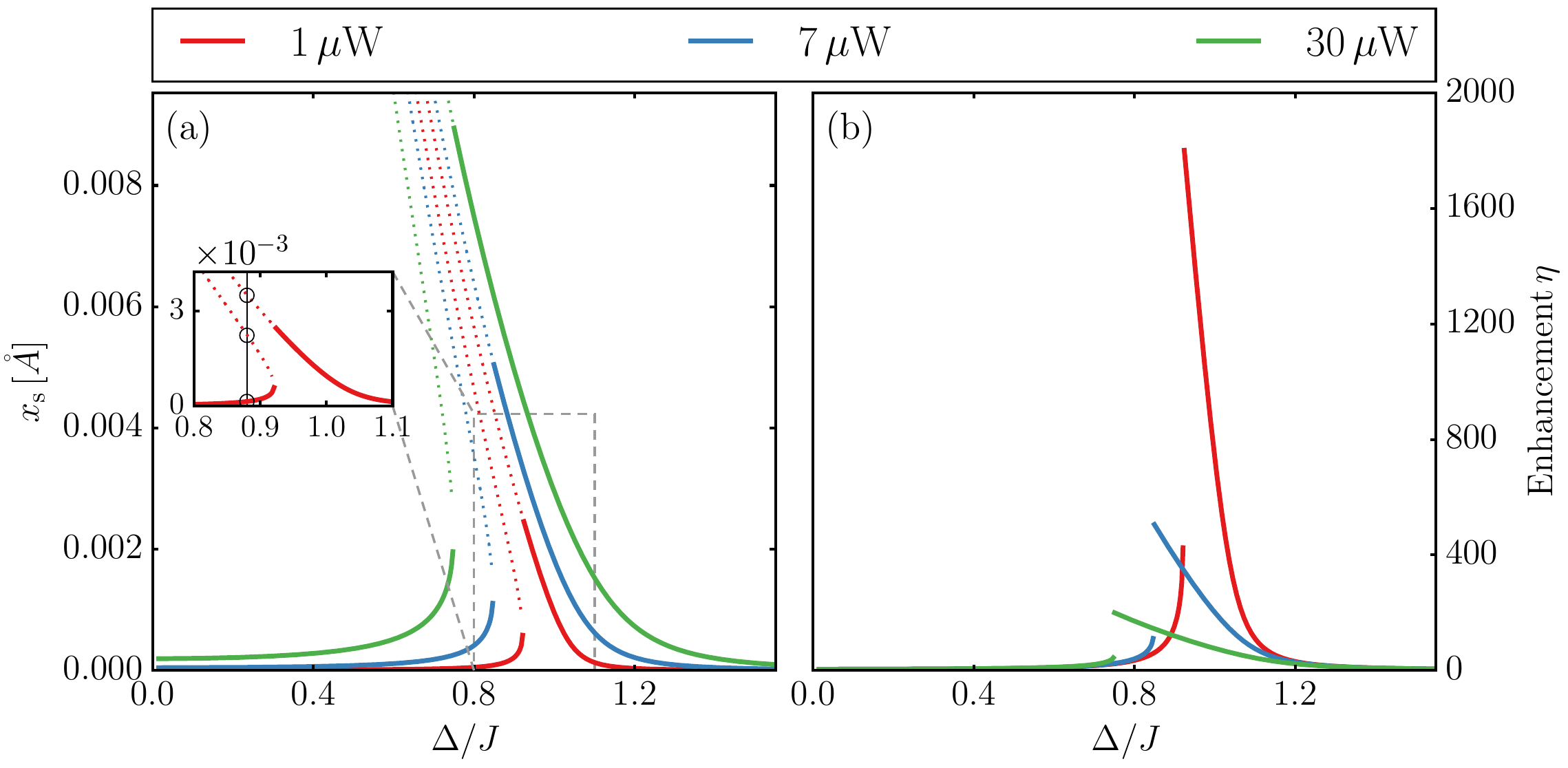}
\caption{Steady-state amplitude $x_\mathrm{s}$ (a) as well as enhancement factor $\eta$ (b) as a function of the detuning $\Delta$ for various laser powers $P_\mathrm{in}$ and $\gamma_1=\gamma_2=0$. The red, blue, and green lines (top to bottom in the right panel, bottom to top in the left panel) correspond to laser powers of $P_\mathrm{in}=1\mu$W, $7\mu$W, and $30\mu$W, respectively. Dotted lines indicate real solutions that diverge as $\Delta\rightarrow 0$. Other parameters as in \Tref{tab:parameters}.}
\label{fig:Fig4_xs_eta_gammakappazero_deltascan}
\end{figure}

Finally, we consider the special case of zero-loss and zero-gain, i.e., $\gamma_1=\gamma_2=0$, shown in \Fref{fig:Fig4_xs_eta_gammakappazero_deltascan}. Under this condition, two different regimes for the steady-state solutions of $x_\mathrm{s}$ can be identified depending the excitation detuning $\Delta$. In particular, within the range $0<\Delta<\Delta_\mathrm{B}$, with $\Delta_\mathrm{B}/J \sim 0.92/0.85/0.75$ for laser drivings of $1/7/30\,\mu$W, three real solutions exist for $x_\mathrm{s}$. ($\Delta_\mathrm{B}$ denotes the branching detuning value, i.e., the detuning at which the single real solution for  $x_\mathrm{s}$ branches into three real solutions.) Note that two of these solutions diverge as $\Delta \rightarrow 0$. Conversely, when $x_\mathrm{s}>\Delta_\mathrm{B}$, only one real solution exists. This is illustrated in \Fref{fig:Fig4_xs_eta_gammakappazero_deltascan}(a) where the diverging branches of the real solutions for $x_\mathrm{s}$ are indicated by dotted lines whereas the finite ones are plotted with solid lines. \Fref{fig:Fig4_xs_eta_gammakappazero_deltascan}(b) shows the enhancement corresponding to the steady-state values of the finite branch in (a) (evaluated with respect to a reference state with $\gamma_1=\gamma_2=-\gamma_0=-J$ as before).

Considering Figs.~\ref{fig:Fig2_eta_kappasmall_deltascan}(b) and \ref{fig:Fig4_xs_eta_gammakappazero_deltascan}(b), we see that the enhancement $\eta$ of the mechanical steady-state amplitude obtained in a loss-gain balanced system as compared to a system with both cavities experiencing equal loss can be outperformed by introducing nonzero detuning. In addition, even in the $\mathcal{PT}$ symmetric case a larger enhancement $\eta$ can be obtained when increasing the inter-cavity coupling $J$ and tuning the laser frequency to the supermode resonance frequency (cf.\ Fig.~\ref{fig:Fig3_eta_J_deltascan}). 

\subsection{Stability analysis of the steady-state solutions}\label{subsec:stability_analysis}
We have so far investigated only steady-state solutions. An important feature of these solutions is their stability. In fact, any steady-state solution is dynamically meaningless unless it is stable. Here we carry out the linear stability analysis of the fixed points of Eqs.~\pref{eq:semiclassical_Langevin_eq} by linearizing Eqs.~\pref{eq:semiclassical_Langevin_eq} around the steady-state values \cite{strogatz2014}. To do so, we start by rewriting Eq.~\pref{eq:semiclassical_Langevin_eq_x} as
\begin{subequations}\label{eq:rewrite_x_ode_first_order}
  \begin{align}
    \dot{x} &= v, \nonumber\\
    \dot{v} &= -\Gamma v -\omega_m^2 x + \frac{\hbar g}{m} |a_2|^2 \nonumber.
  \end{align}
\end{subequations}
By introducing a perturbation vector $\delta\vec{q} = (\delta a_{1r},\delta a_{1i},\delta a_{2r},\delta a_{2i},\delta x, \delta v)^T$ (cf.\ also Ref.~\cite{lue2015}) over any particular steady-state solution, substituting back in Eqs.~\pref{eq:semiclassical_Langevin_eq} and neglecting higher order terms, we find $\delta\dot{\vec{q}}=\bsb{M}\delta\vec{q}$, where the matrix $\bsb{M}$ is given by
\begin{equation}\label{eq:coeff_matrix}
 \bsb{M} = 
    \begin{pmatrix}
      \gamma_1 & \Delta & 0 & J & 0 & 0\\
      -\Delta & \gamma_1 & -J & 0 & 0 & 0\\
      0 & J & \gamma_2 & g x_\mathrm{s} +\Delta & g a_{2i,\mathrm{s}} & 0\\
      -J & 0 & -g x_s -\Delta & \gamma_2 & -g a_{2r,\mathrm{s}} & 0\\
      0 & 0 & 0 & 0 & 0 & 1 \\
      0 & 0 & 2\hbar g a_{2r,\mathrm{s}}/m & 2\hbar g a_{2i,\mathrm{s}}/m & -\omega_m^2 & -\Gamma
    \end{pmatrix}.
\end{equation}
The matrix $\bsb{M}$ is the Jacobian matrix associated with perturbations of the steady state of the nonlinear system of Eqs.~\pref{eq:semiclassical_Langevin_eq}; the subscript $\mathrm{s}$ denotes steady state and $r,i$ denote real and imaginary parts, respectively, of the amplitudes $a_{1,2}$. The stability of steady-state solutions for any set of given design/excitation parameters depend on the eigenvalues of $\bsb{M}$. In particular, a given steady-state solution is stable if all eigenvalues of $\bsb{M}$ have negative real parts (note that $\bsb{M}$ is a function of the steady-state solutions and varies from one to another). In that case, this solution is represented by a fixed point surrounded by an attracting region in phase space, meaning that all trajectories in the vicinity of this fixed point will converge into it. Otherwise, if some of the eigenvalues have positive real parts, the steady state becomes \emph{unstable} and might exhibit limiting cycles or display chaotic behavior \cite{strogatz2014,lue2015}.

\begin{figure}[tb]
\centering
\includegraphics[width=.8\columnwidth]{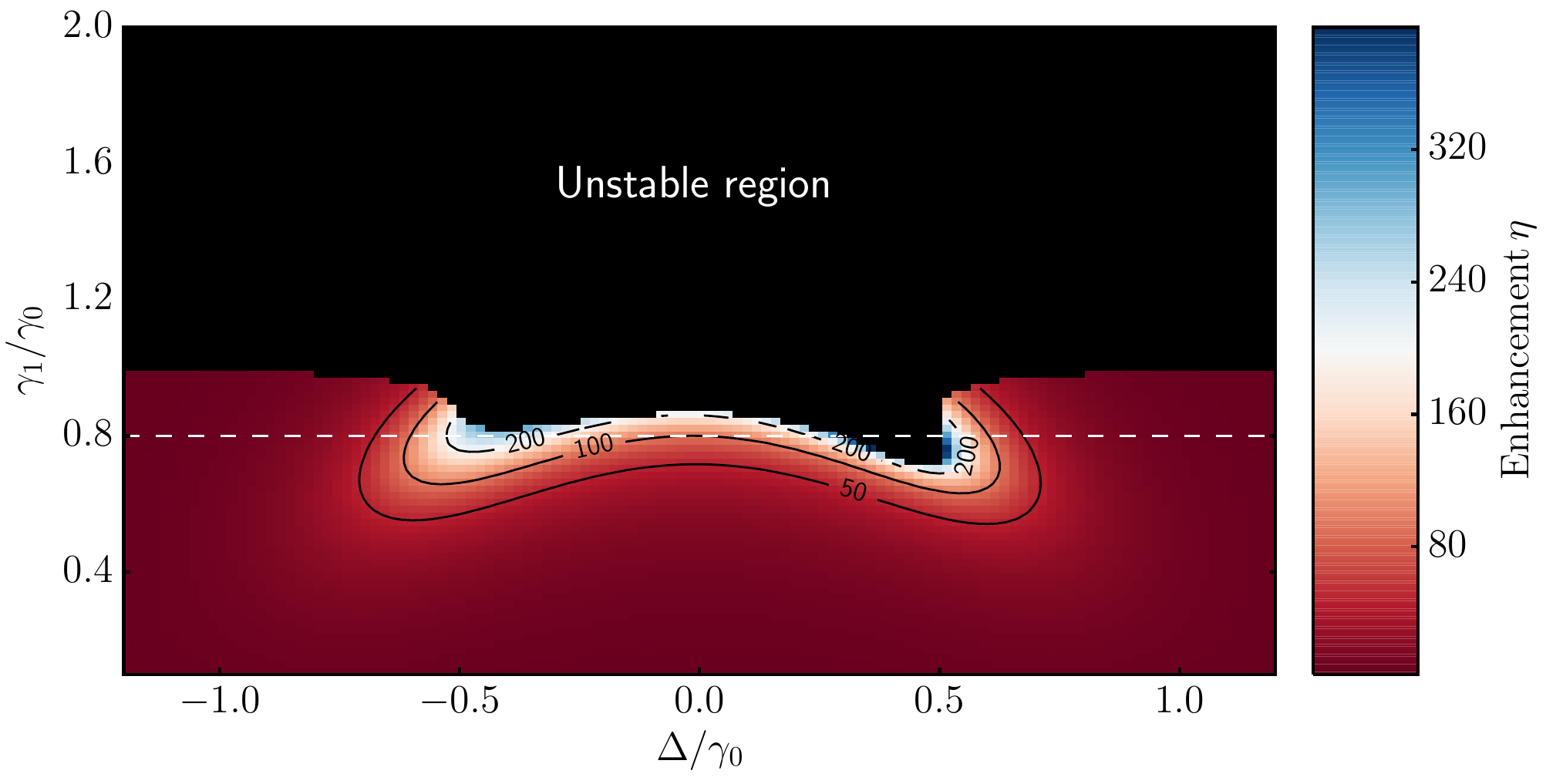}
\caption{Enhancement $\eta$ as a function of the detuning $\Delta$ and the gain-to-loss ratio $\gamma_1/\gamma_2\equiv\gamma_1/\gamma_0$ at a driving power of $P_\mathrm{in}=1$ $\mu$W. All other parameters are shown in \Tref{tab:parameters}. The black region indicates parameter regimes where steady-state solutions are not stable according to linear stability analysis. Contours of equal enhancement are also shown. The horizontal white dashed line corresponds to the parameters of \Fref{fig:Fig2_eta_kappasmall_deltascan}(b).}
\label{fig:Fig5_eta_kappaoverdelta_stability}
\end{figure}

By constructing a linear stability map for the fixed points of Eqs.~\pref{eq:semiclassical_Langevin_eq} as a function of the gain $\gamma_1>0$ and detuning $\Delta$ parameters (see \Fref{fig:Fig5_eta_kappaoverdelta_stability}), we uncover the following remarkable result: Steady-state solutions that correspond to the $\mathcal{PT}$ symmetric case $\gamma_1=-\gamma_2=\gamma_0$ are not stable. In other words, $\mathcal{PT}$ symmetry is not necessarily the optimal choice for enhancing optomechanical interactions at steady state. Instead, \Fref{fig:Fig5_eta_kappaoverdelta_stability} shows that stable steady-state solutions that exhibit significant enhancement (up to $200$ fold) can be still achieved for nonzero pump detuning and broken $\mathcal{PT}$ symmetry. In particular, the gain values must satisfy $\gamma_1/\gamma_0\lesssim 0.7$ in order to guarantee stability over the full range of the considered detuning. Hence, the peak enhancement in the case of \Fref{fig:Fig2_eta_kappasmall_deltascan}(b), indicated by the white dashed line in \Fref{fig:Fig5_eta_kappaoverdelta_stability}, as well as that reported in \cite{jing2014} is indeed misleading since it does not correspond to stable steady-state solutions. As we will show later, including gain saturation effects can result in a stable steady-state solution even in the $\mathcal{PT}$ symmetric case.

We conclude this section by noting that while linear stability analysis suffices to question the validity of claims made on the basis of steady-state analysis alone, it does not provide much information regarding the dynamical behavior of the system and whether it converges to a limit cycle or even becomes chaotic. In order to explore the full behavior of the system, we numerically integrate the full temporal dynamics associated with Eqs.~\pref{eq:semiclassical_Langevin_eq}.

\section{Nonlinear dynamics}\label{sec:dynamics}
In the previous section, we studied the stability properties of steady-state solutions associated with optomechanical photonic molecules having optical gain and loss profiles (see \Fref{fig:Fig1_system}). We have shown that in the case of $\mathcal{PT}$ symmetry (equal gain and loss) the phase space fixed points are unstable. We also revealed that steady-state solutions that exhibit significant enhancement in optomechanical interactions can be attained by tailoring the pump detuning and the gain/loss profile (with unbalanced distribution). This analysis however leaves several important questions unanswered: 
(1) What are the dynamics when the steady-state solutions are unstable? (2) What is the effect of gain saturation?

\begin{figure}[tb]
\centering
\includegraphics[width=.8\columnwidth]{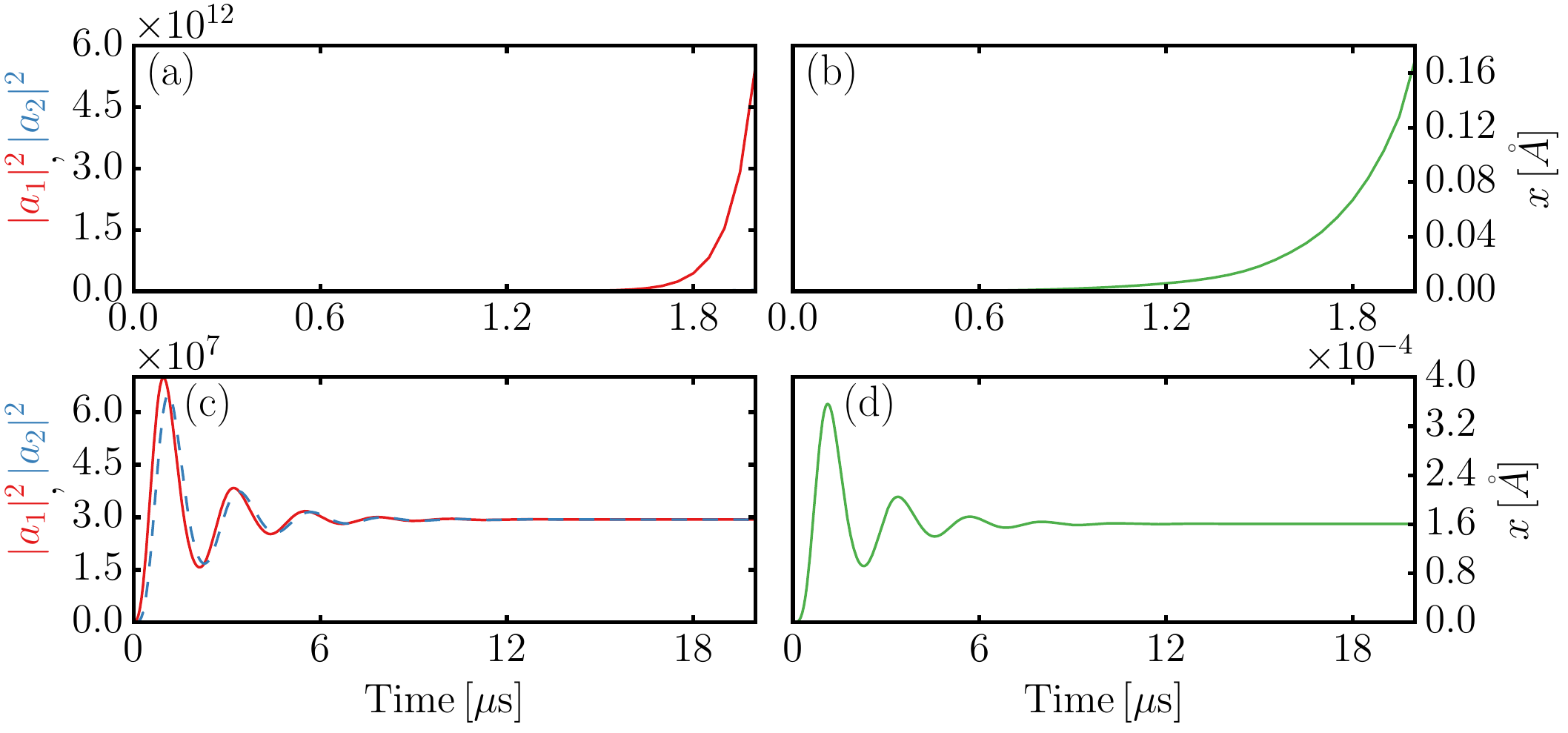}
\caption{Dynamics of the populations $|a_1(t)|^2$ (left column, solid red), $|a_2(t)|^2$ (left column, dashed blue) and the mechanical oscillator amplitude $x(t)$ (right column, solid green). In (a) and (b), $\gamma_1=-\gamma_2=\gamma_0$ whereas in the (c),(d) $\gamma_2=-\gamma_0$ and $\gamma_1=0.8\gamma_0$. The laser power is $P_\mathrm{in}=1\mu$W and $\Delta=0$ MHz; all other parameters as in \Tref{tab:parameters}. Note the different scaling of the $x$ and $y$ axes.}
\label{fig:Fig6_dynamics_divergent_steadystate_comparison}
\end{figure}

In this section we investigate the above posed questions. To do so, we begin by studying the temporal evolution of the dynamical quantities $|a_1(t)|^2, |a_2(t)|^2$ and $x(t)$ for the two different cases depicted in \Fref{fig:Fig2_eta_kappasmall_deltascan} ($\mathcal{PT}$ symmetry and unbalanced gain and loss, respectively) when the detuning is zero and for an input laser power of $P_\mathrm{in}=1\mu$W.  By integrating Eqs.~\pref{eq:semiclassical_Langevin_eq} numerically, we find that, in the first case of $\mathcal{PT}$ symmetric gain and loss distribution where $\gamma_1=-\gamma_2=\gamma_0$,  the optical intensities and mechanical displacement grow exponentially as shown in Figs.~\ref{fig:Fig6_dynamics_divergent_steadystate_comparison}(a) and (b) (Note that we do not study long-time dynamics subsequent to the exponential growth, which might exhibit chaotic features \cite{lue2015}.) In contrast, Figs.~\ref{fig:Fig6_dynamics_divergent_steadystate_comparison}(c) and (d) show that for unbalanced gain and loss, $\gamma_1=0.8\gamma_0$ and $\gamma_2=-\gamma_0$, the steady state is reached on a timescale of $\sim 10\mu$s.

While these results are consistent with stability analysis, it is important to note that in general, the unbounded exponential growth in the first case cannot continue indefinitely. In fact, gain saturation mechanisms \cite{kepesidis2015} are expected to regulate these divergences. 

In particular, a full model should include a gain coefficient of the form $\gamma_1/(1+|a_1|^2/a_s^2)$ \cite{chang2014,kepesidis2015} with $a_s$ being the gain saturation threshold, rather than just $\gamma_1$. By taking this effect into account, we find that the divergent mechanical oscillation amplitude behavior in Fig.~\ref{fig:Fig6_dynamics_divergent_steadystate_comparison}(b) indeed reaches a steady-state value.
In contrast to our previous finding in the case of unsaturated gain, for appropriate gain saturation threshold we now obtain steady-state solutions even in the $\mathcal{PT}$ symmetric case ($\gamma_2=-\gamma_1$ and $\Delta=0$), with enhancement factors ranging from $\eta\sim 8$ for $a_s=10^3$ to $\eta\sim 340$ for $a_s=3\times10^4$ (other paramters are $\gamma_1=J=\gamma_0$ and $P_\mathrm{in}=1\mu$W).

\begin{figure}[tb]
\centering
\includegraphics[width=.8\columnwidth]{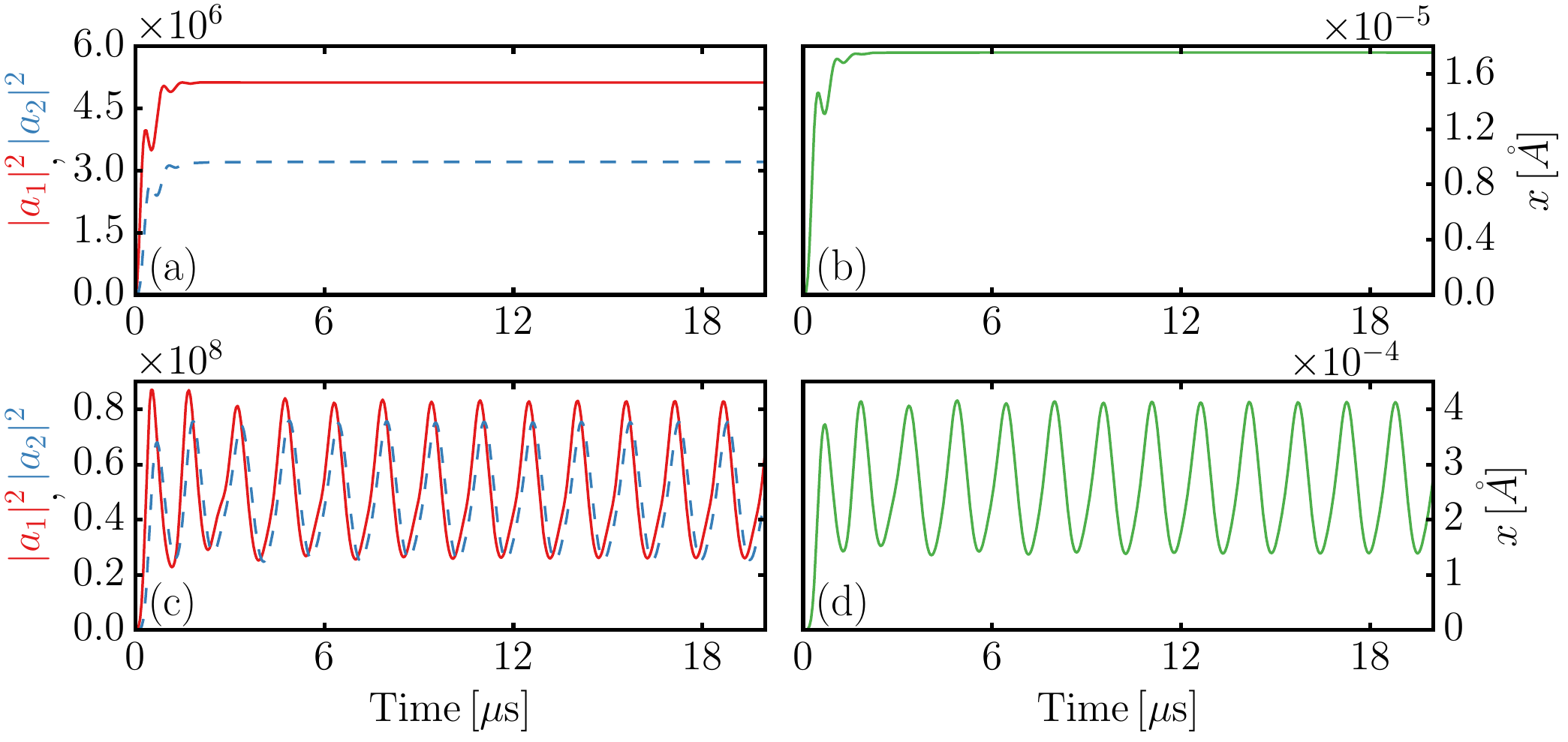}
\caption{Dynamics of the populations $|a_1|^2$ (left column, solid red), $|a_2|^2$ (left column, dashed blue) and the mechanical oscillator amplitude $x(t)$ (right column, solid green). In (a) and (b), the saturation parameter $a_s=10^3$ whereas in (c),(d) $a_s = 10^4$. In both plots, $\gamma_2=-\gamma_0$, $\gamma_1=1.5 \gamma_0$, $\Delta = -5$ MHz and $P_\mathrm{in}=1\mu$W; all other parameters as in \Tref{tab:parameters}. Note the different scaling of the $y$ axes.}
\label{fig:Fig7_dynamics_as_saturation_comparison}
\vspace{-5pt}
\end{figure}

Moreover, when we choose the parameters such that $\gamma_2=-\gamma_0$, $\gamma_1=1.5 \gamma_0$, $\Delta = -5$MHz and $P_\mathrm{in}=1\mu$W, we find two different dynamical regimes depending on the value of the gain saturation. In particular, as shown in \Fref{fig:Fig7_dynamics_as_saturation_comparison}, whereas the system reaches a steady state when $a_s = 10^3$, the dynamics converges to a sustained oscillation reminiscent of an oscillator limit cycle for $a_s = 10^4$. This feature might illustrate the importance of accounting for gain saturation effects in order to understand reported mechanical oscillatory dynamics \cite{grudinin2010}.

\begin{figure}[tb]
\centering
\includegraphics[width=.8\columnwidth]{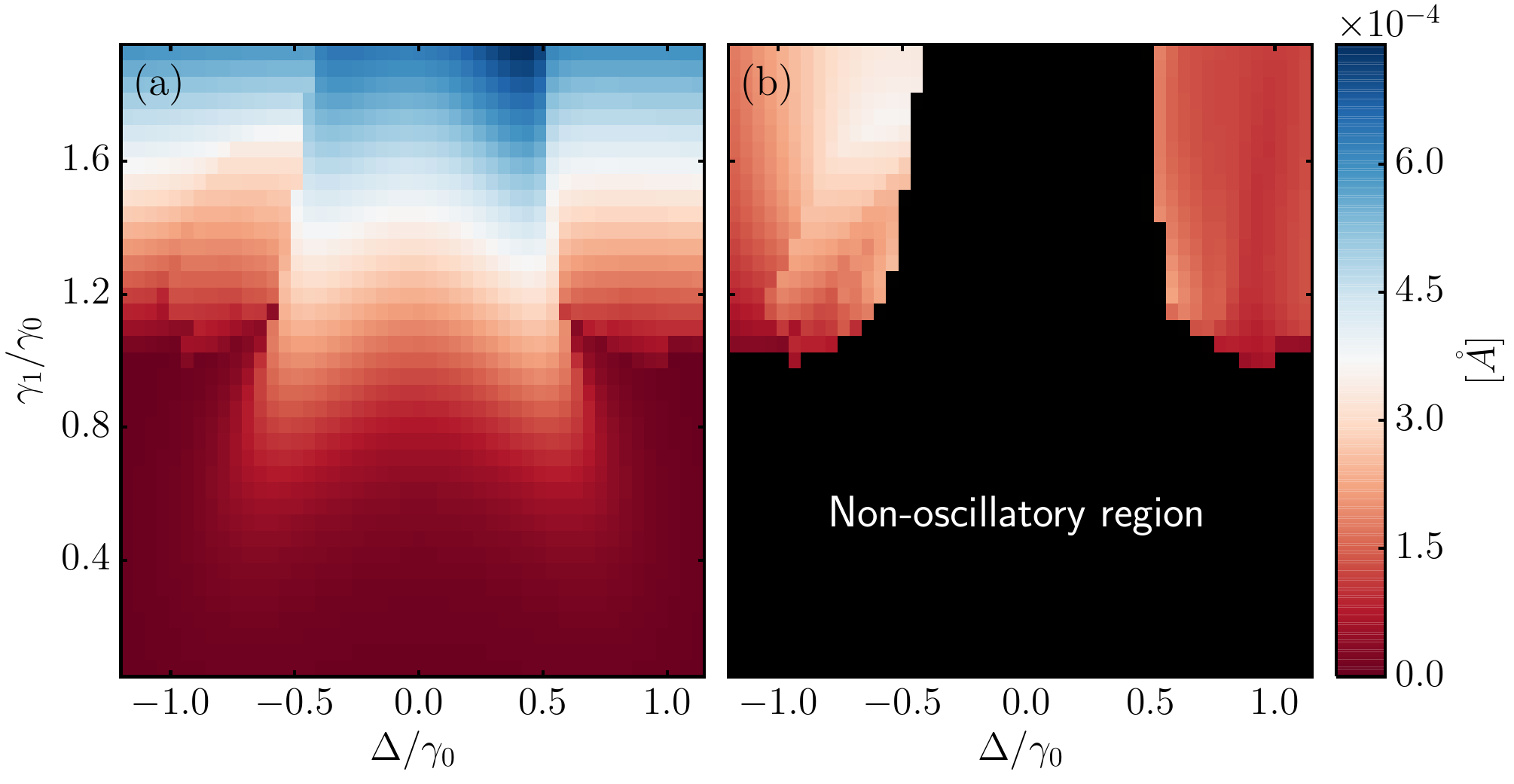}
\caption{Analysis of the dynamics of the mechanical amplitude in the presence of gain saturation, $a_s=10^4$. In (a), the mean of the oscillation amplitude $\langle x\rangle$ is shown whereas (b) shows the amplitude of the oscillation (maximal amplitude minus minimal amplitude). Back shading in (b) indicates parameter regimes where a steady state rather than an oscillatory motion is reached. The laser power $P_\mathrm{in}=1\mu$W; all other parameters as in \Tref{tab:parameters}.}
\label{fig:Fig8_dynamics_saturation_gamma1overDelta}
\vspace{-5pt}
\end{figure}

Finally, in order to gain more insight into the dynamics of the mechanical degree of freedom in the presence of an effective gain with $a_s=10^4$, we evaluate the mean values as well as oscillation amplitudes as a function of gain and detuning in \Fref{fig:Fig8_dynamics_saturation_gamma1overDelta}. Interestingly, near resonant pumping $\Delta=0$, the mechanical oscillator always relaxes towards a steady state. On the other hand, oscillatory behavior can occur for off-resonant driving; thus highlighting the rich dynamics associated with these systems under different conditions. 

\section{Conclusion and discussion}\label{sec:conclusions}
In conclusion, we have carried out a comprehensive study of the static and dynamic behavior of optomechanical interaction in non-Hermitian photonic molecules that support an acoustic mode. 
Our steady-state analysis demonstrates that the strength of the interaction between the photonic supermodes and the mechanical oscillators of an active (gain/gain) system as compared to a passive (loss/loss) system can be significantly enhanced under different conditions for design and pump parameters. Interestingly, we found that $\mathcal{PT}$ symmetry is not necessarily the optimal choice for achieving maximum enhancement (compared with the passive system). Instead, we have shown that pump frequency detuning can lead to higher enhancement values. 

Furthermore, we have studied the linear stability properties of these systems and we have shown that the enhancement factors reported in the $\mathcal{PT}$ symmetric case near the exceptional point correspond to unstable solutions. In this regard, we have identified regions in parameter space that correspond to unbalanced optical gain/loss distribution and laser detuning where much stronger interactions (two orders stronger than the passive cavities) can be still achieved for linearly stable solutions. In addition, we have also investigated the dynamical evolution of the system by numerically integrating the nonlinear equations. Our analysis revealed that gain saturation effects play an important role in regulating the behavior of the otherwise exponentially growing oscillations that correspond to unstable fixed points. Moreover, two distinct dynamical behaviors were identified based on the physical and pump parameters: stable fixed points and sustained oscillations.

It is worth noting that in this work we have focused on the classical aspects of non-Hermitian photonic molecules that can exhibit exceptional points of order two \cite{teimourpour2014} and demonstrated their rich behavior. It would be of interest to investigate the behavior of similar optomechanical systems in photonic networks having higher order exceptional points \cite{teimourpour2014}. Another interesting aspect is to explore the quantum properties of these systems. While some quantum aspects were discussed briefly in \cite{jing2014}, proper treatment using left/right eigenvalue algebra of non-Hermitian Hamiltonians is still lacking. We plan to carry out these investigations elsewhere.

\section*{Acknowledgments}
R E would like to acknowledge the support from Elizabeth and Richard Henes Center for Quantum Phenomena at Michigan Technological University.

\appendix
\section{Detailed discussion of the efficiency $\eta$ in the two-resonator model}\label{sec:appendix}
In this appendix we discuss the steady-state solutions of Eqs.~\pref{eq:semiclassical_Langevin_eq_illustration} in more detail. Including laser detuning $\Delta$, the enhancement $\eta = |a_2|^2 / |a_{2,\mathrm{p}}|^2$ reads as
\begin{equation}\label{eq:eta_illustration_full}
  \eta = \frac{J^4 + 2 J^2 (\gamma_2^2 - \Delta (\Delta + \Delta_x)) + (\gamma_2^2 + \Delta^2) (\gamma_2^2 + (\Delta + \Delta_x)^2)}{J^4 + 2 J^2 (\gamma_1 \gamma_2 - \Delta (\Delta + \Delta_x)) + (\gamma_1^2 + \Delta^2) (\gamma_2^2 + (\Delta + \Delta_x)^2)}.
\end{equation}
From this equation, two limits are readily obtained. That is, for large detuning $\Delta$, the enhancement $\eta$ goes as
\begin{equation}\label{eq:illustration_limit_delta}
 \eta \overset{|\Delta|\rightarrow \infty}{\longrightarrow} 1,
\end{equation}
in agreement with \Fref{fig:Fig2_eta_kappasmall_deltascan} and \Fref{fig:Fig5_eta_kappaoverdelta_stability}. On the $\mathcal{PT}$ point ($\gamma_2=-\gamma_1=-J<0$),
\begin{equation}
 \eta = 1+\frac{4\gamma_2^2}{\Delta_x^2},
\end{equation}
which is Eq.~\pref{eq:eta_estimate_illustration} in the main text. Note that letting $\Delta_x\rightarrow\infty$ independent of, e.g., the detuning $\Delta$ is misleading because the mechanical steady-state displacement $x_\mathrm{s}$ does exhibit a detuning dependence in the full model. However, the qualitative behavior of the enhancement with transmitted laser power is captured even in the simple model, which is why we employ it for instructive purposes.

\section*{References}
\providecommand{\newblock}{}

\end{document}